\documentclass{aa} 
\input{epsf} 
\usepackage{graphicx}
\begin{document}

\title{XMM-Newton observation of the Tycho Supernova Remnant 
\thanks{Based on an observation obtained with XMM-Newton, an ESA science 
mission with instruments and contributions directly funded by ESA Member 
States and the USA (NASA).} }
 
\author{A.~Decourchelle \inst{1}, J.L.~Sauvageot \inst{1}, M.~Audard 
\inst{2}, B.~Aschenbach \inst{3}, S.~Sembay \inst{4}, 
R.~Rothenflug \inst{1},  J.~Ballet \inst{1}, T.~Stadlbauer \inst{3},
R.G.~West \inst{4} } 
   \authorrunning{Decourchelle et al.}
   \offprints{A.~Decourchelle}

   \institute{Service d'Astrophysique, CEA Saclay, F-91191 Gif-sur-Yvette 
Cedex, France
   \and Paul Scherrer Institute, Laboratory for Astrophysics, CH-5232 
Villigen PSI, Switzerland
   \and MPI f\"ur extraterrestrische Physik, 85741 Garching, Germany
   \and Department of Physics and Astronomy, Leicester University, 
Leicester LE1 7RH, United Kingdom} 
 
   \date{Received October 6, 2000; accepted November 6, 2000} 
 
\abstract{
 We present the observation of the Tycho supernova remnant obtained 
with the EPIC and RGS instruments onboard the {\it XMM-Newton} satellite.
We compare images and azimuthally averaged radial profiles in emission lines 
from different elements (silicon and iron) and different transition lines of iron 
(Fe L and Fe K). While the \ion{Fe}{xvii} L line and \ion{Si}{xiii} K line 
images are globally spatially coincident, the Fe K emission clearly peaks at 
a smaller radius, indicating a higher temperature toward the reverse shock. 
This is qualitatively the profile expected when the reverse shock, after 
travelling through the outer power--law density profile, has entered the 
central plateau of the ejecta. The high energy continuum map has an overall 
smooth distribution, with a similar extent to the radio emission. Its radial 
profile peaks further out than the lines emission. 
Brighter and harder continuum regions are observed with a rough 
bipolar symmetry in the eastern and western edges.
The spectral analysis of the southeastern knots supports spatial variations 
of the relative abundance of silicon and iron, which implies an incomplete 
mixing of the silicon and iron layers.
} 
\maketitle 
%
 
\section{Introduction} 

The Tycho supernova remnant (SNR) is considered as the prototype for Type Ia 
remnants (Baade 1945). The supernova was recorded in 1572 by Tycho Brahe, 
hence, the remnant is relatively young and still retains much information 
about the progenitor. Its X-ray emission is dominated by shocked ejecta.
With an angular diameter of 8 arcmin, Tycho SNR is well suited for investigating 
the spatial distribution of heavy elements in the ejecta. Hwang \& Gotthelf 
(1997) presented deconvolved line images 
of the object using {\it ASCA} data. They found that the Fe K radial 
profile peaks at smaller radius than the other emission lines. This was 
interpreted as a possible iron stratification in the interior of the ejecta.

With {\it XMM-Newton} observatory, it is now possible to perform true 
spatially resolved spectroscopy of Tycho SNR, allowing the mapping of the 
emission for individual elements (and individual ions) and for the continuum 
at high energy.
In this paper we present a first study of Tycho SNR with {\it XMM-Newton}.

 
\section {Observation}

Tycho SNR was observed by {\it XMM-Newton} (Jansen et al. 2001) on 2000 June 29 
with the two X-ray instruments EPIC (European Photon Imaging Camera; Str\"uder 
et al. 2001; Turner et al. 2001) and RGS (Reflection Grating Spectrometer; den 
Herder et al. 2001). After selecting good time intervals (when the light curve in 
background regions is stable, i.e. uncontaminated by flaring events), the 
available exposure time was 12 ks for each EPIC MOS camera and 15 ks for the 
EPIC PN camera. Due to the high signal-to-noise (S/N) ratio of the RGS spectra of 
Tycho and the smaller sensitivity of the RGS instruments to background variations, 
no time-selection was applied to the RGS data, resulting in a total of 45 ks of 
exposure time. The X-ray image of Tycho is entirely contained within 
the central CCD of each of the MOS cameras whereas in the PN camera it is 
spread over 6 CCDs, resulting in data gaps between each CCD. For simplicity, 
therefore, we have only used the EPIC MOS data in our spatial analysis.

The three EPIC spectra (MOS~1, MOS~2 and PN), integrated over the entire 
remnant, are shown in Fig.~1, with the RGS spectrum obtained on a 
region smaller than Tycho (the FOV is shown in Fig.~2). The most prominent 
lines in the EPIC spectra are the He~${\alpha}$ lines of 
magnesium, silicon, sulphur, argon, calcium, iron and an \ion{Fe}{xvii} line 
in the Fe L complex. With the RGS, the remnant has been detected up to the 
5th order, however with a low S/N ratio from the third order. 
Emission lines from \ion{Si}{xiii}~He~${\alpha}$, \ion{Mg}{xi}~He~${\alpha}$, 
\ion{Ne}{x}~Ly~${\alpha}$, \ion{Ne}{ix}~He~${\alpha}$, two \ion{Fe}{xvii} 
lines (at 15 \AA~and 17 \AA) and \ion{O}{viii}~Ly~${\alpha}$ have been 
identified. The structure seen both in EPIC and RGS spectra between the magnesium 
and neon lines (10~\AA~and 12~\AA) is attributed to Fe L shell lines.  Both 
RGS data sets were analyzed and are fully consistent with each other. 

\section{Elemental structure in the ejecta} 

The bright emission lines of heavy elements (silicon, sulphur, iron) in 
the spectrum of the entire remnant originate from shocked material within 
the supernova ejecta. 

We have created images selecting photons in narrow energy bands 
corresponding to the main line features. For the maps of \ion{Si}{xiii} and 
\ion{Fe}{xvii}, we did not substract the continuum: these two lines have the 
highest S/N ratio above the continuum. For the lower S/N ratio Fe K$\alpha$ 
line, where the spectral shape is steep, the continuum does need to be 
subtracted. This was done by using the overall global spectrum to derive the 
ratio of the continuum at the line energy to the continuum in an adjacent 
energy band, free of contamination by line emission. 
In each spatial bin, the continuum at the line was estimated using the flux in 
the adjacent energy band and the previously defined ratio.

To complement this image approach, we present azimuthally averaged radial 
profiles of the deconvolved line images. For the MOS1 and MOS2 cameras, the 
FWHM of the point spread function is ~4.4\arcsec~at 1500~eV and is slightly 
larger at higher energy (Aschenbach et al. 2000, Gondoin et al. 2000). 

We have used a standard Van Cittert deconvolution algorithm (Starck et al. 
1998) to reconstruct the images within typically 20 iterations. 

\begin{figure}
\begin{center}
\resizebox{7.cm}{!}{\includegraphics[angle=-90]{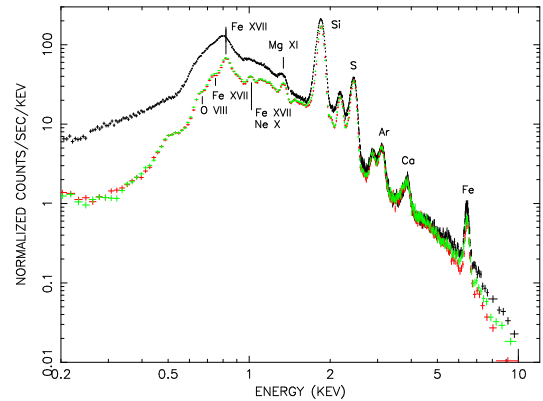}}
\resizebox{8.cm}{9.cm}{\includegraphics{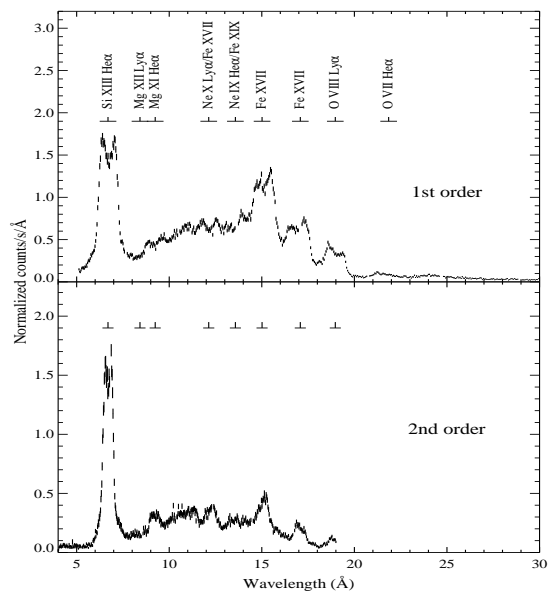}}
\caption{Top: EPIC spectra integrated over the whole remnant obtained with the 
MOS 1 (green), MOS 2 (red) and PN (black) camera. Bottom: RGS 1 spectrum of a 
limited region of Tycho (see Fig.~2) with order 1 and order 2. The spatial 
extent of the SNR dominates the line profiles.}
\end{center}
\label{spectot}
\end{figure}

\subsection{Large scale structure}

The 15~\AA~\ion{Fe}{xvii} line image is shown in Fig.~2, 
overlaid with the \ion{Si}{xiii} He~$\alpha$ contours. Overall, the 
spatial distribution is very similar, although Fig.~2 shows that the image, 
as well as the radial profile, in the \ion{Si}{xiii} line is probably 
slightly peaked exterior to the \ion{Fe}{xvii} line image. The good 
correlation between the emission from iron and silicon implies that the 
outer layers of the supernova have been at least partially mixed. This is 
predicted by numerical hydrodynamical modeling of the global X-ray spectrum 
of Tycho (Itoh et al. 1988; Brinkmann et al. 1989).

\begin{figure}
\begin{center}
\resizebox{7.cm}{!}{\includegraphics{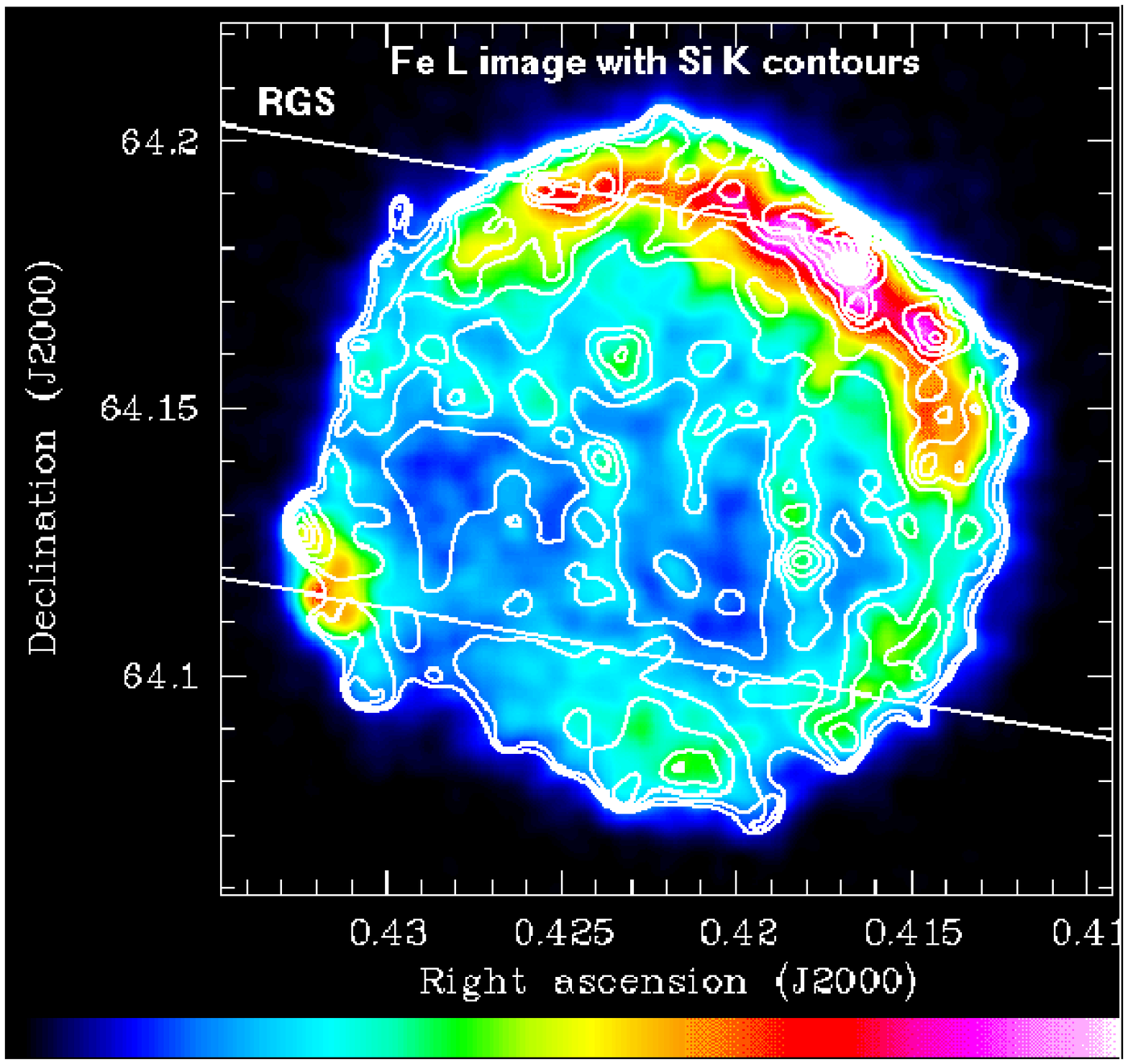}}
\resizebox{6.cm}{!}{\includegraphics[angle=+90]{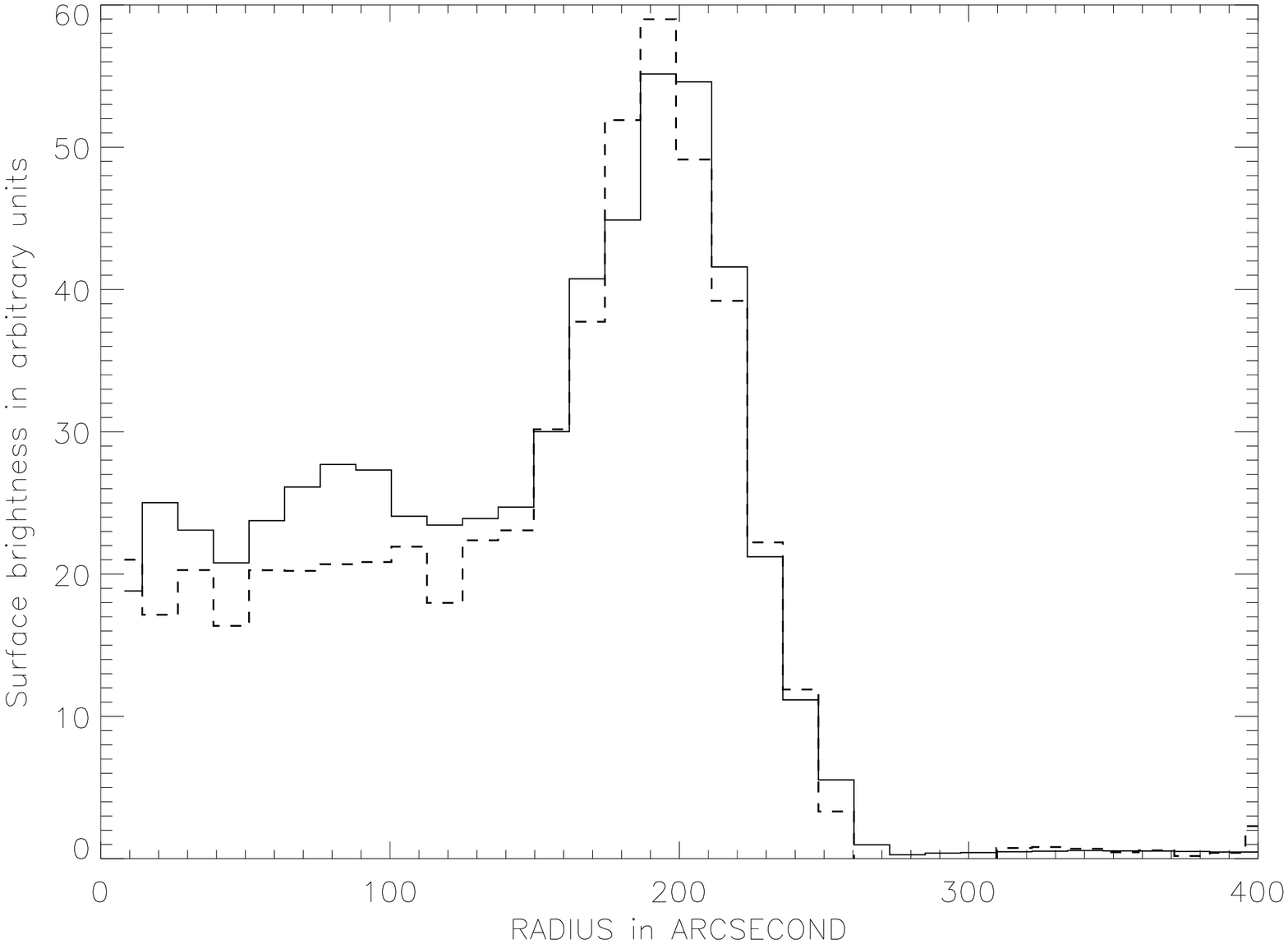}}
\caption{Top: Contours of the \ion{Si}{xiii} triplet blend (1670-2000 eV) 
superimposed on the EPIC MOS image in the \ion{Fe}{xvii} ($\lambda=15$~\AA, 
775-855 eV) line. The RGS field of view is overlaid and the dispersion axis 
is parallel to the solid white lines. The Right Ascension is in decimal 
degrees.
Bottom: Comparison of the azimuthally averaged radial profiles of the 
deconvolved images in the \ion{Si}{K} (solid line) and \ion{Fe}{xvii} lines 
(dashed line).}
\end{center}
\label{Felsik}
\end{figure}

\begin{figure}
\begin{center}
\resizebox{7.cm}{!}{\includegraphics{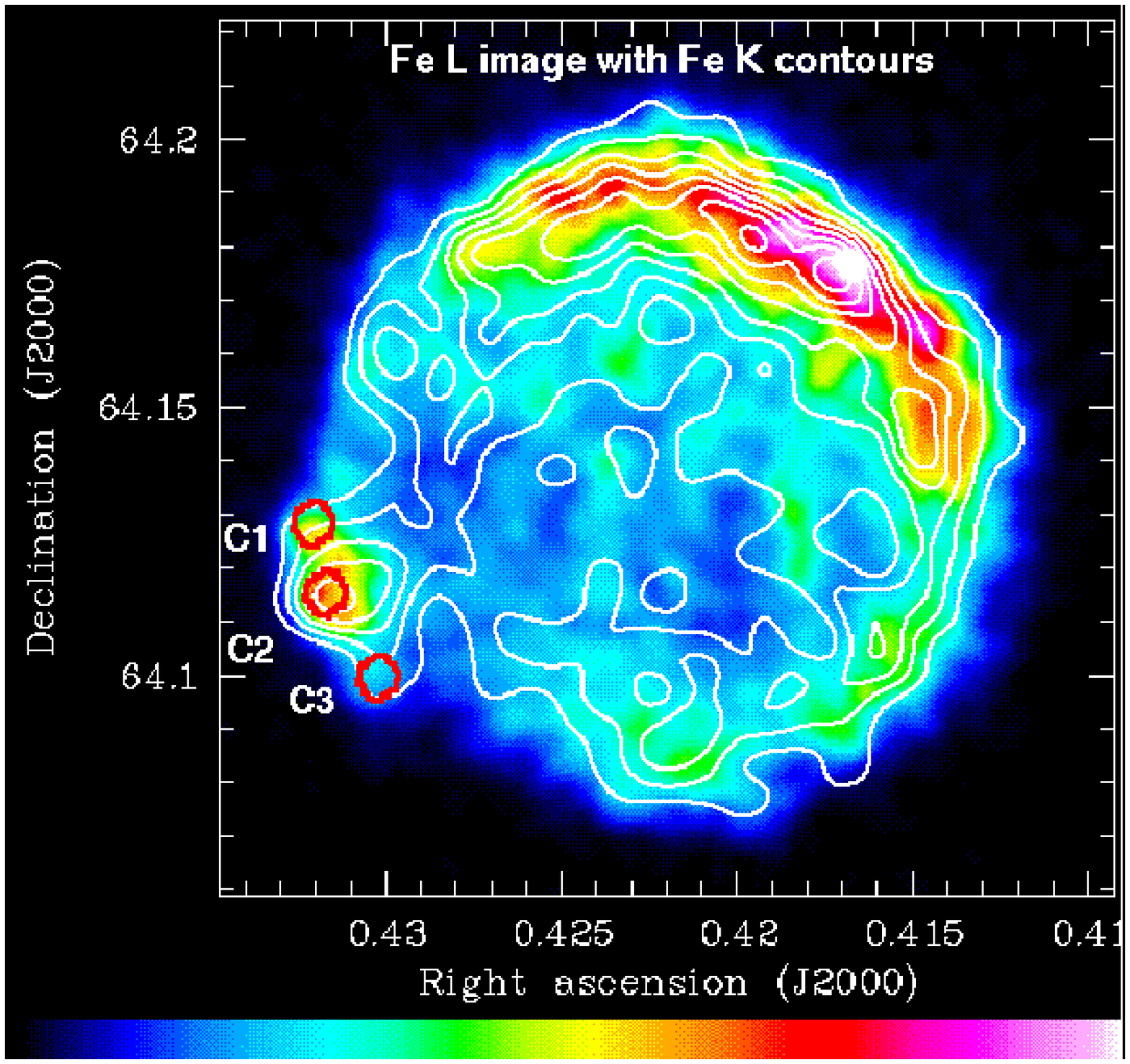}}
\resizebox{6.cm}{!}{\includegraphics[angle=+90]{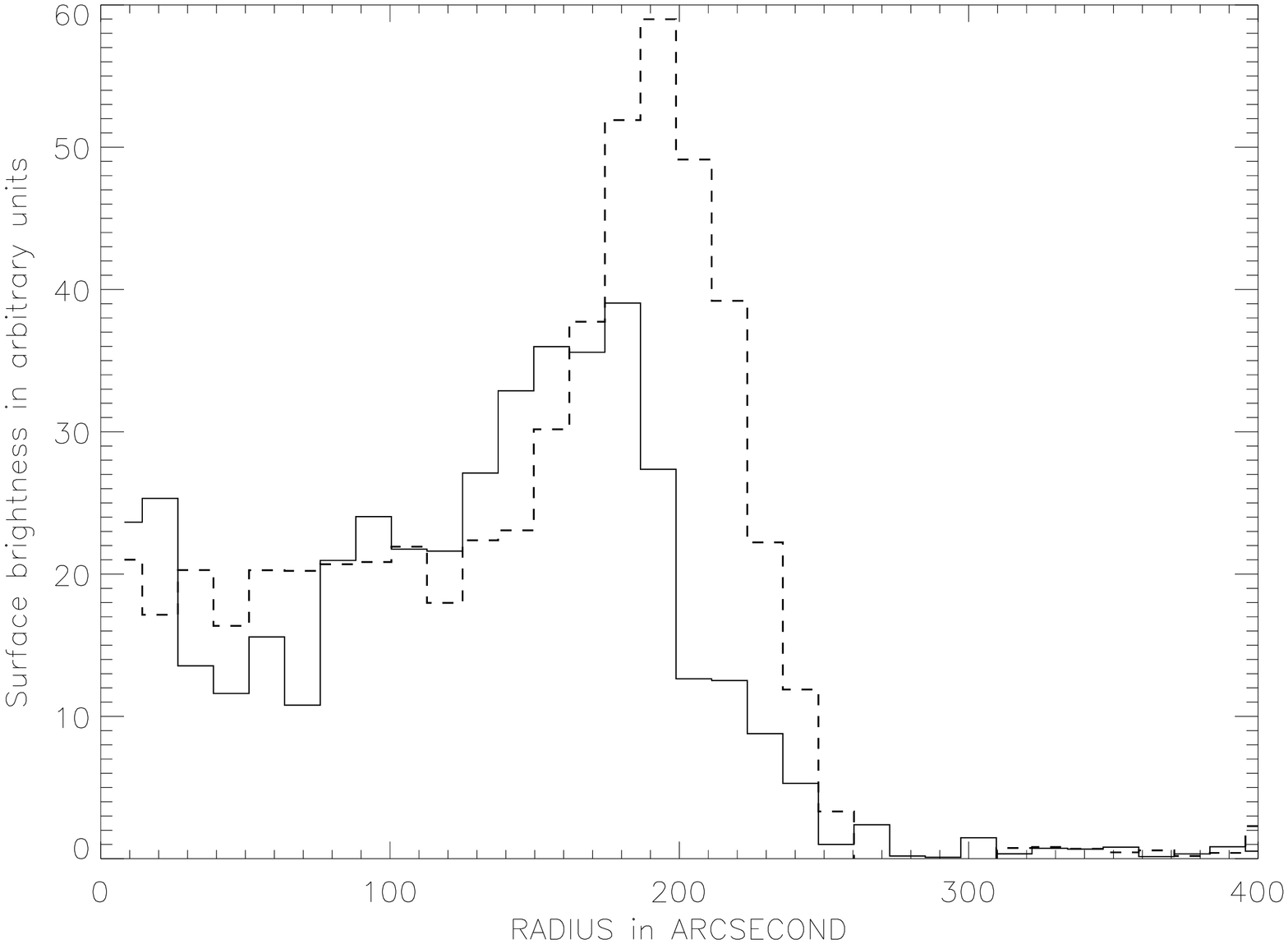}}
\caption{Top: Contours of the Fe K line (6200-6600 eV) image superimposed on 
the EPIC MOS image in the \ion{Fe}{xvii} line. Fields C1, C2 and C3 are shown 
with red solid lines.
Bottom: Comparison of the radial profiles of the deconvolved images in the 
Fe K (solid line) and \ion{Fe}{xvii} lines (dashed line).}
\end{center}
\label{FeLFeK}
\end{figure}

In Fig.~3, we show again the 15~\AA~\ion{Fe}{xvii} line image, overlaid 
with the contours in the continuum subtracted Fe K$\alpha$ line image. 
There is an overall agreement between the line images, with spherical 
emission brightest in the northwest, faintest in the east and intermediate 
in the south sectors as for \ion{Si}{xiii}. Both Fe lines exhibit strong 
emission in the southern part of the southeast knots, and a faint contribution 
to the northern part for \ion{Fe}{xvii}. The comparison of the radial profiles 
and of the images demonstrates that the spatial distribution of the Fe K$\alpha$ 
emission is broader, and peaks at smaller radius, than the \ion{Fe}{xvii} 
emission, as was first suggested by Hwang and Gotthelf (1997), based on 
{\it ASCA} deconvolved narrow band images.

The fact that the \ion{Si}{xiii} He $\alpha$ and \ion{Fe}{xvii} images 
correlate well overall, whilst the Fe L and Fe K lines of iron are clearly 
displaced, favors the interpretation of a spatial structure in the 
temperature and ionization timescale in the shocked ejecta. This is indeed 
what is expected from hydrodynamical evolution models of young supernova 
remnants (see for example Chevalier 1982 and Decourchelle \& Ballet 
1994). Although a spatial segregation of the elements is possible, 
depending on the degree of mixing, a structure in temperature and/or 
ionization stage is in most cases unavoidable and depends on the initial 
density profiles of the supernova.

The Fe K line corresponds to higher temperature material (at least twice 
higher) and lower ionization timescale than the Fe L and silicon emission 
(Hwang et al. 1998). This indicates that the temperature in the 
inner shocked ejecta is higher with a lower ionization age than the outer 
shocked ejecta region. Such profiles of temperature and ionization age are 
those qualitatively expected from the evolution of a supernova remnant 
of 1.4~M$_\odot$ in a uniform interstellar medium, when the reverse shock 
is propagating, from an outer power--law density profile, into the inner 
plateau region of the ejecta. For an explosion energy of 10$^{51}$ ergs 
and an ambient density of 0.35~amu~cm$^{-3}$, the reverse shock, at the age 
of Tycho, has already entered the central plateau (Decourchelle 1994). The 
dense shocked ejecta region (colder and older), 
close to the contact discontinuity, is the result of the interaction of a 
power--law density ejecta profile with the ISM (Chevalier 1982) at the 
beginning of the evolution. When the reverse shock reaches the central 
plateau, the density begins to decrease in the ejecta, the temperature 
increases and the X-ray emissivity diminishes (see for example Fig.~6 in 
Band \& Liang 1988). This is qualitatively what we observe in Tycho with 
the inner Fe K line in a region with fainter emission. 

Standard models of the Type Ia supernova explosion, such as the W7 model 
(Nomoto et al. 1984), give outer density profiles in the 
ejecta closer to an exponential decline than to a power law. 
However, the characteristic temperature profiles for exponential ejecta 
expanding into a constant density ambient medium are flatter than
the temperature gradient required in Tycho, unless a circumstellar medium is 
invoked (Dwarkadas \& Chevalier 1998). Note however that for stratified 
ejecta, the heavier element layers would have a higher temperature because of 
the lower number of electrons for the same mass density.

\subsection{Small scale inhomogeneities}

A more careful inspection of the comparison between the silicon and iron 
images (see Fig.~2) shows that of the southeastern knots (Vancura 
et al. 1995) the northern one is brightest in the \ion{Si}{xiii} line, whereas 
the southern one is brightest in the \ion{Fe}{xvii} line.  

\begin{figure}
\begin{center}
\resizebox{7.cm}{!}{\includegraphics[angle=-90]{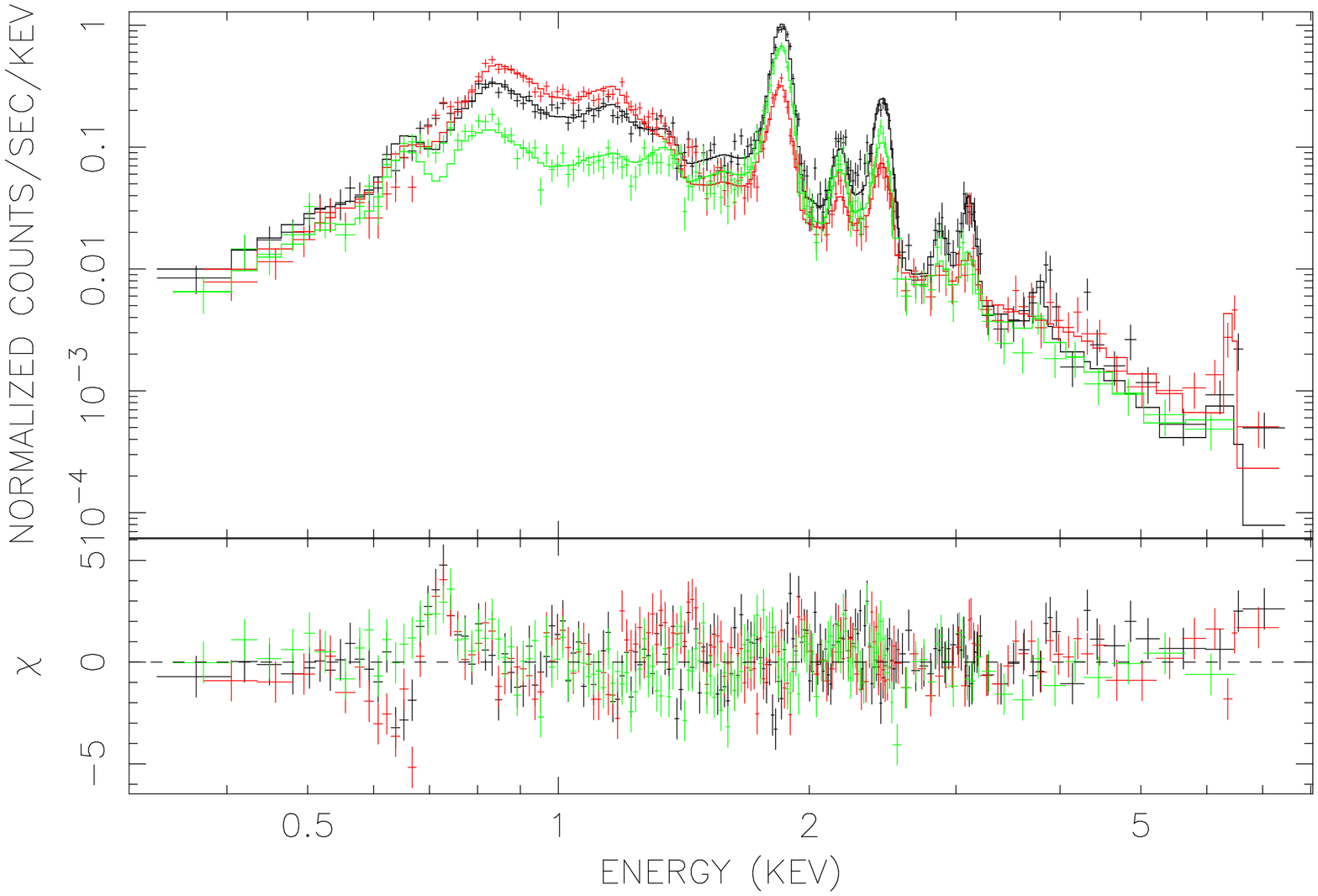}}
\resizebox{7.cm}{!}{\includegraphics[angle=-90]{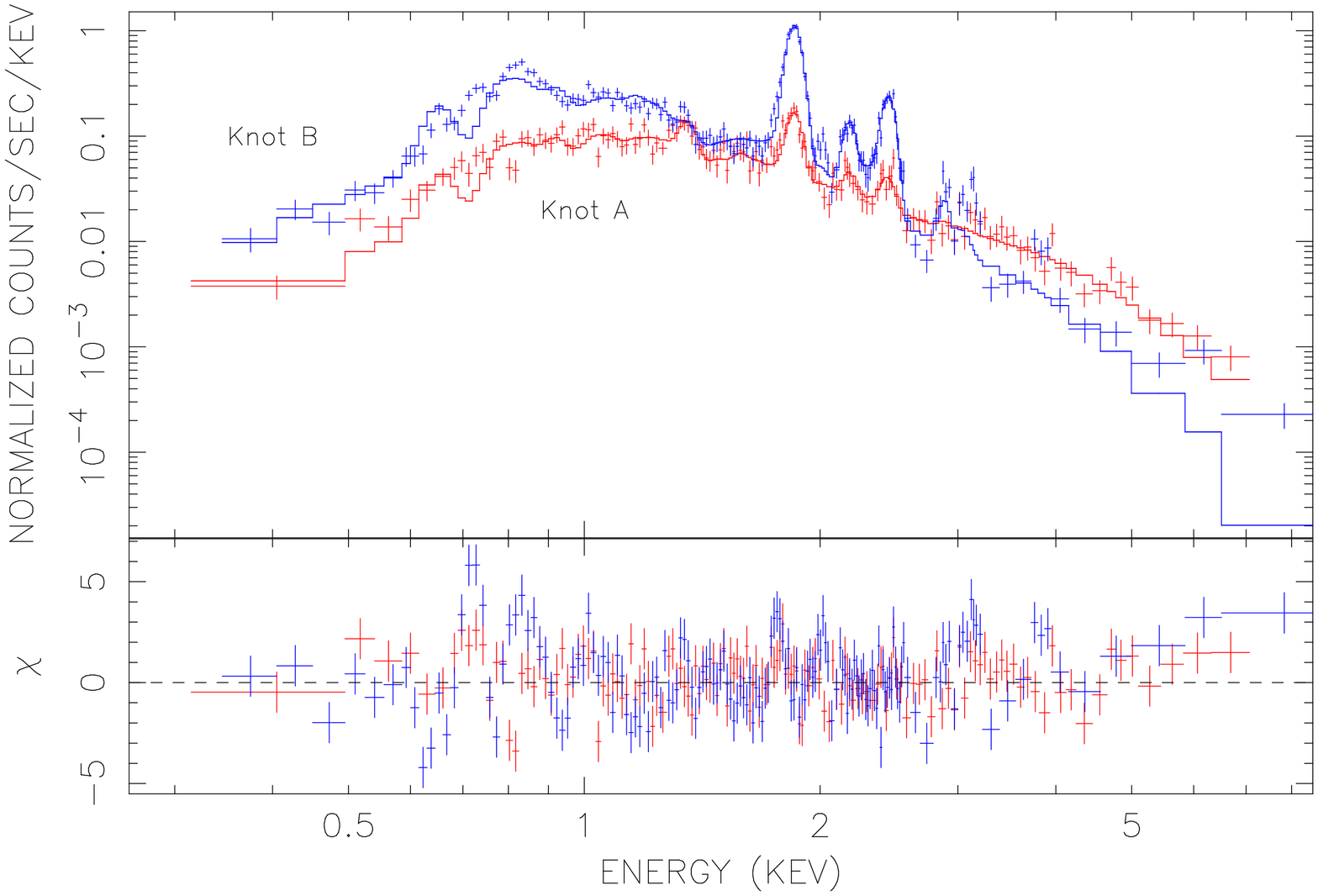}}
\caption{Top: MOS 1 and MOS2 Spectra of the three southeastern knots C1 (red), 
C2 (black) and C3 (green). Bottom: Spectra of a strong continuum region 
(Knot A in red) and a line dominated region (Knot B in blue). The regions A 
and B are shown in Fig.~5.}
\end{center}
\label{knotsC}
\end{figure}

Fig.~4 (top panel) shows the MOS~1 and MOS~2 spectra of the three southeastern 
knots, marked C1, C2 and C3 on Fig.~3.
We have extracted spectra from these identically sized regions and fit them 
with a two component non-equilibrium model (with variable abundances for the first 
component). A Gaussian line at 3.1 keV was 
added to the model to account for the argon emission line, which is missing 
in the non-equilibrium model VNEI within XSPEC (Arnaud 1996). 

The first component is at a relatively low temperature ($T_1$) and accounts 
for the line emission which is the focus of this discussion, whilst the 
second component assumes solar abundances, is at a high temperature and 
accounts for the high energy continuum. We assume no hydrogen and helium in 
the shocked SN Ia ejecta, and solar values for carbon and nitrogen. 

The three different spectra can be simultaneously fitted with the same temperature 
(k$T_1=0.62 \pm 0.02$ keV), ionization timescale 
($\rm 1.2 \times 10^{11}~s~cm^{-3}$) and hydrogen column density 
($\rm 0.83 \times 10^{22}~cm^{-2}$) and a relatively good fit is achieved 
($\chi^2 = 1183/721$ dof). However, there are, between the knots, clear 
variations of the abundances of silicon, sulphur and iron, which are given in 
Table 1. 

\begin{table}
\begin{flushleft}
\begin{center}
\caption[ ]{Derived abundances (relative to the solar photospheric values) 
in the knots}
\begin{tabular}{llllll}
\hline
   & C1 & C2 & C3  & A & B \\
\hline
Si & 2.1 & 1.1 & 1.6  & 1.0 & 4.2 \\
S  & 4.6 & 2.2 & 2.1  & 1.1 & 6.9 \\
Fe & 0.2 & 0.4 & 0.06 & 0.1 & 0.7 \\
\end{tabular}
\end{center}
\end{flushleft}
\end{table}

The spectral analysis demonstrates that the different emission structures 
seen in the southeastern knots are primarily due to spatial variations of the 
relative abundances of silicon, sulphur and iron, rather than temperature and 
ionization timescale effects. The relative abundance of sulphur to silicon 
is roughly constant in these three knots, while that of iron is varying. 
Knot C1 has the largest silicon and sulphur abundance, whereas knot C3 
has almost no iron. In this particular knot C3, almost a factor of 10 less 
iron has protruded into the silicon layer compared with the adjacent knot 
C2. However, the southeastern knots are 
the only place where such a relative variation of silicon to iron is 
unambigously observed. As shown in Fig.~2 and discussed in section 3, there is 
otherwise a good agreement between the images in \ion{Fe}{xvii} and in 
\ion{Si}{xiii} for the whole remnant. These knots resemble to some extent 
the silicon rich shrapnels seen in Vela (Aschenbach et al. 1995; 
Tsunemi et al. 1999).

\section{High-energy continuum and radio emission}  

The high-energy continuum image in the 4500-5800~eV band is shown in Fig.~5, 
overlaid with the \ion{Si}{xiii} emission contours. The image in the 
continuum is more regular and extended than in 
the silicon line, which shows irregular outer contours, which can be 
attributed to Rayleigh-Taylor instabilities at the contact discontinuity 
between the ejecta material and the ambient medium (see Fig.~6).
 The extent of the continuum emission corresponds well with the radio VLA 
image at 22 cm (Dickel et al. 1991) and gives the position 
of the outer blast wave. This is shown in Fig.~5, where the radio image 
is overlaid with the contours of the continuum emission. The regular spherical 
emission of the continuum, unlike the emission line images, may indicate that 
the ambient medium surrounding Tycho was fairly homogeneous, 
although it has been recently interacting on its eastern edge with \ion{H}{i} 
clouds (Reynoso et al. 1999). The more asymmetric shape in the line emission 
may be related to azimuthal variations of the heavy element distribution in 
the ejecta.

The radial profiles of the X-ray continuum and radio emission are shown in 
the bottom panel of Fig.~5. The peak of the continuum emission is the 
outermost and is interpreted as emission from the shocked interstellar 
medium. The radio emission extends as far as the continuum but the profile 
peaks at about the same radius as the Si emission (Fig.~2). For comparison the 
profile of the Fe K emission is also shown and clearly peaks at a much
smaller radius than the other components.

\begin{figure}
\begin{center}
\resizebox{7.cm}{!}{\includegraphics{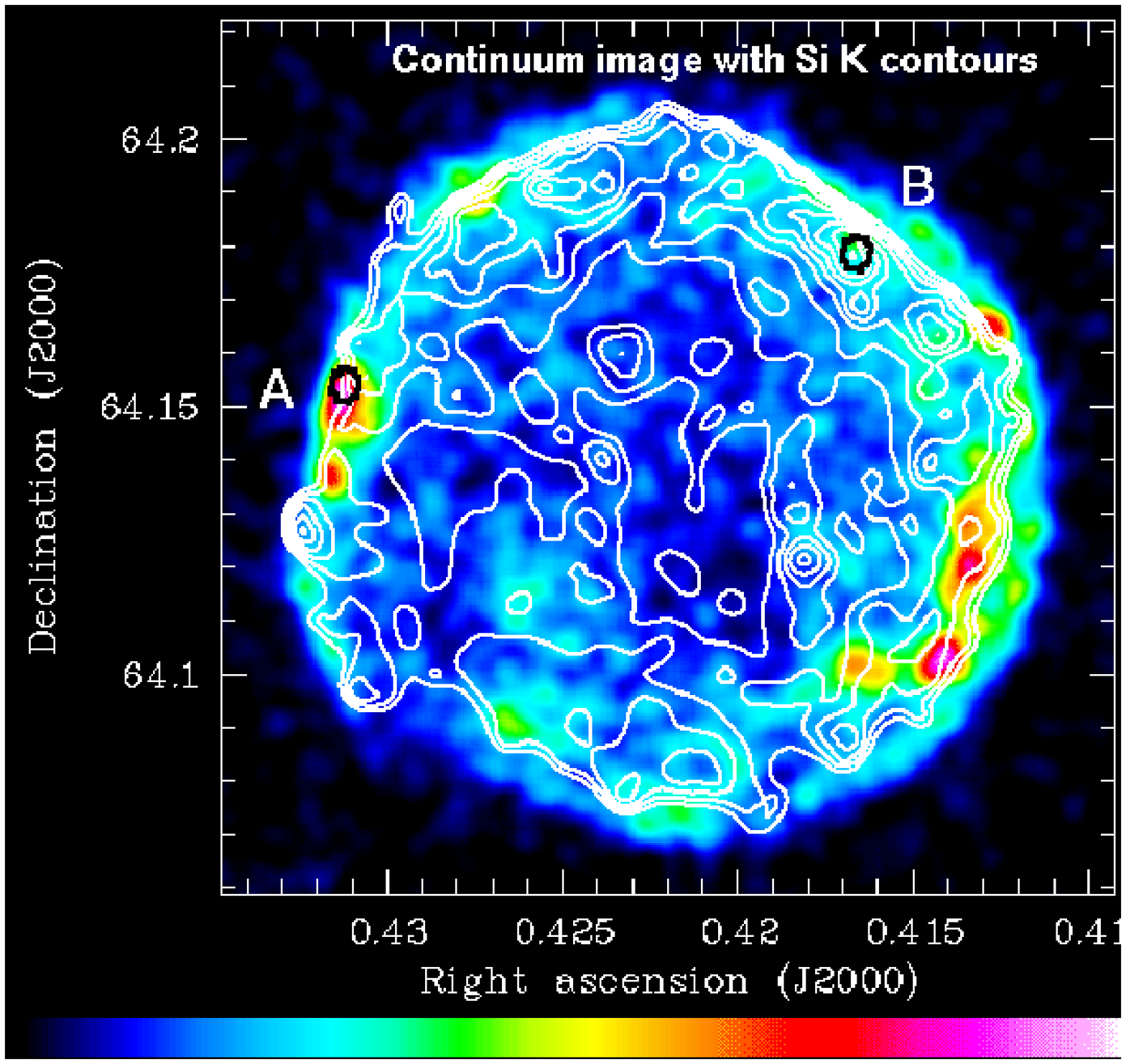}}
\resizebox{7.cm}{!}{\includegraphics{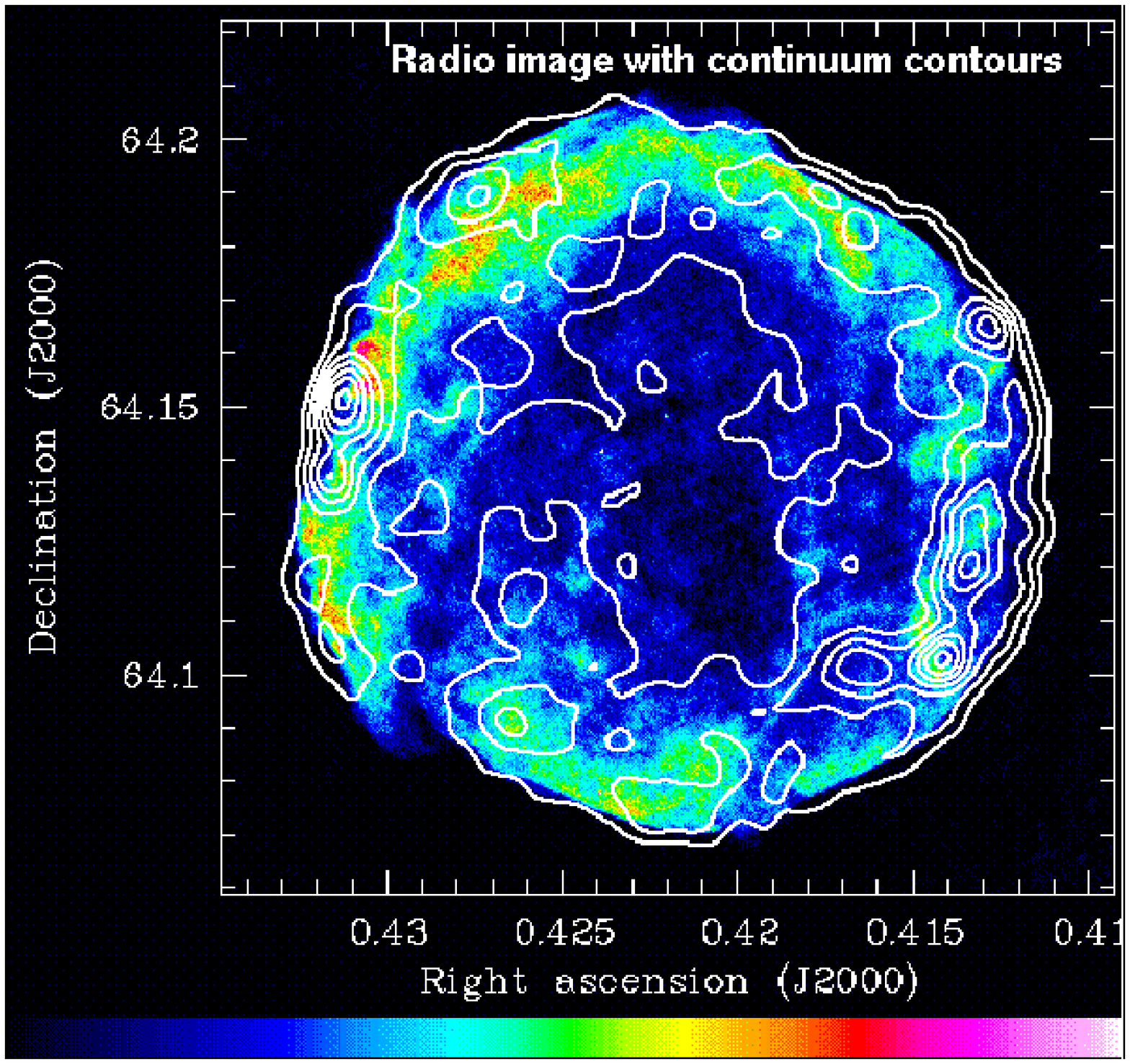}}
\resizebox{6.cm}{!}{\includegraphics[angle=+90]{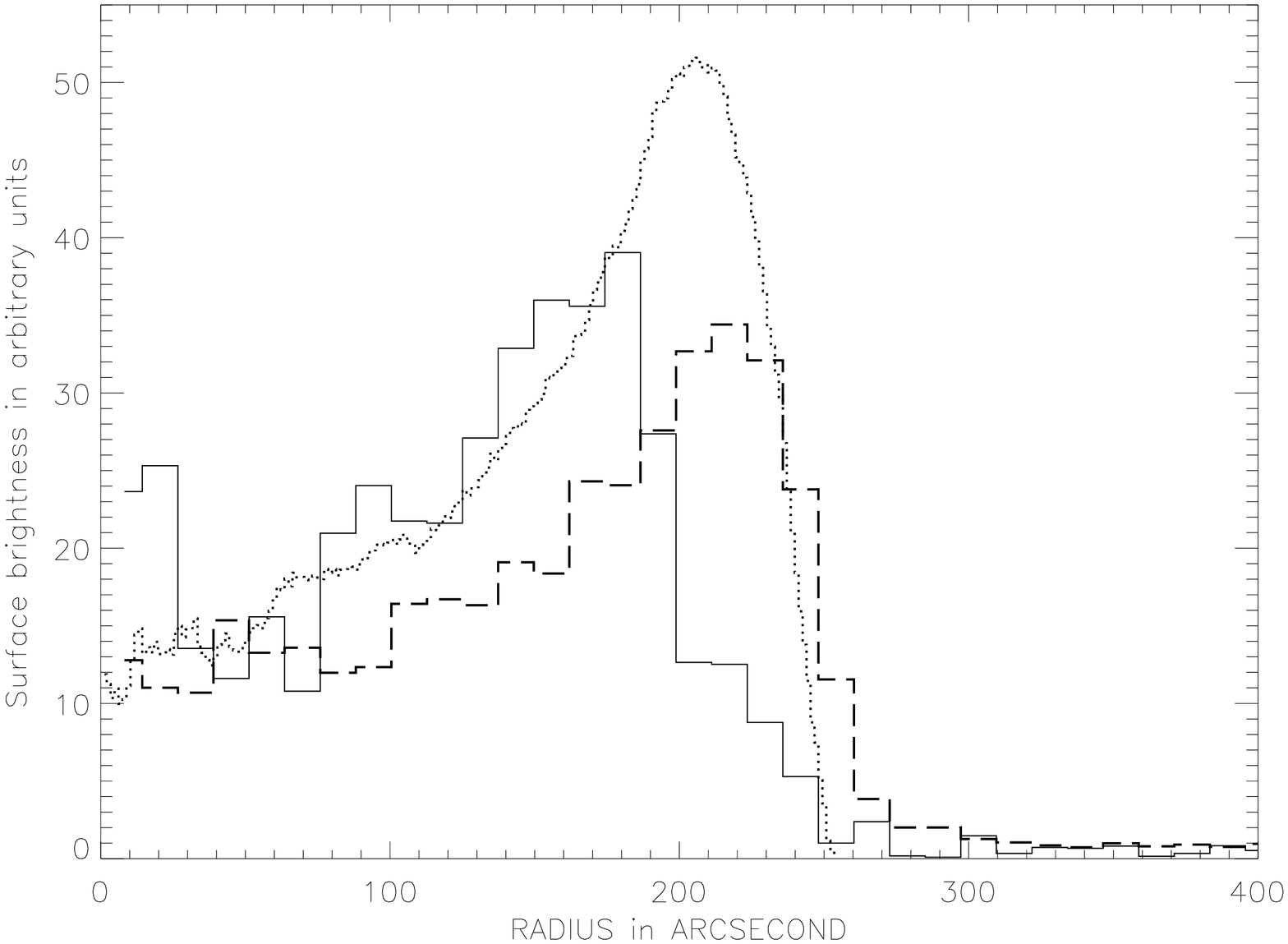}}
\caption{Top: High energy continuum image (4500-5800 eV) overlaid with 
\ion{Si}{xiii} contours. 
Fields A and B are shown with black solid lines. 
Middle: VLA 22 cm radio map overlaid with high energy continuum contours. 
Bottom: Comparison of the three corresponding radial profiles: radio (dotted), 
high energy continuum (dashed) and Fe K line (solid line).}
\end{center}
\label{ContSiK}
\end{figure}

In addition to a regular smooth profile, the continuum map shows striking 
bright knots on the eastern edge, as well as on the diametrically opposite 
side of the remnant. These features are lying inner to the continuum 
edge, they have no clear counterpart in the line images, nor in the radio. 
However, the brightest continuum knot (eastern edge) lies very close to 
the two brightest radio peaks, which exhibit the flattest spectral index 
of the remnant (Filament I, in Katz-Stone et al. 2000), while the southwestern 
continuum knots lie along a radio structure, which also has a flat spectral 
index (Filament A, in Katz-Stone et al. 2000). The eastern knot has the 
lowest radio expansion in Tycho (Reynoso et al. 1997, see Hughes 2000 for the 
X-ray expansion rate), and is close to one of the 
brightest optical filaments (Kamper \& Van den Bergh 1978; Smith et al. 1991), 
associated with infrared emission at 10.7-12 micron (Douvion et al. 2000). A 
recent \ion{H}{i} 21 cm study of the environs of Tycho has found a strong 
\ion{H}{i} absorption feature, corresponding to a cloud of 
density 160-325 cm$^{-3}$ on the eastern edge (Reynoso et al. 1999).

The bright continuum knot lies at the periphery of the line emission image, 
whilst being well inside the continuum emission. The comparison of the 
deconvolved silicon image with the radio image shows that the eastern side is 
the only place where the radio and silicon boundaries overlap and where 
such a strong X-ray continuum is seen (see cross mark in Fig.6). The two 
brightest radio knots coincide with similar features in the silicon image. 
This indicates that they are related to ejecta material, 
with structures characteristic of Rayleigh-Taylor instabilities as argued 
by  Velasquez et al. (1998), based on VLA radio images with a resolution of 
1\arcsec.
%
\begin{figure}
\begin{center}
\resizebox{7.cm}{!}{\includegraphics{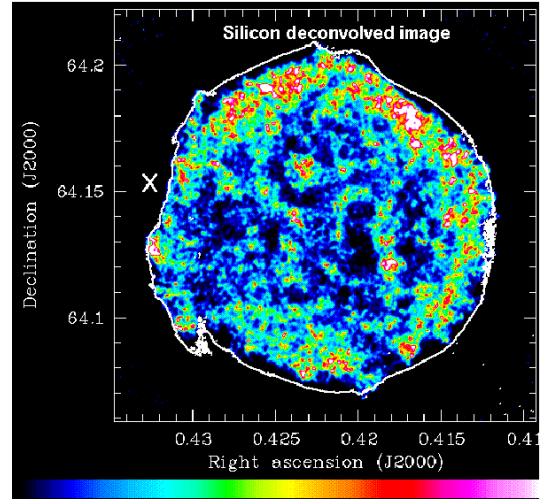}}
\caption{Deconvolved silicon image overlaid with outer contour of the radio 
image from Dickel et al. (1991).}
\end{center}
\label{radioSidec}
\end{figure}

Jun \& Norman (1996) have shown that Rayleigh-Taylor instabilities amplify 
strongly the ambient local magnetic field around the dense features of the 
instability by stretching, winding and compressing the magnetic field lines. 
Assuming an external magnetic field tangential to the shock front, their
simulations account for the observed radial vector polarization observed in 
the remnant (Dickel et al. 1991; Reynoso et al. 1997) and predict an 
amplification of the magnetic field in front of such Rayleigh-Taylor features, 
which would explain the fact that these structures are seen both in radio 
(amplified synchrotron emission around these instabilities) and in X-rays 
(thermal emission of the dense knots).

The strong continuum knot might be the result of the interaction of the 
eastern edge with H~{\small I} clouds, leading either to higher densities and 
magnetic field strengths (hence stronger nonthermal emission), or to a 
reflected shock in front of the cloud, which heats and compresses a second 
time the ISM (hence leading to an increase in the thermal bremsstrahlung 
emission at high energy). However, no \ion{H}{i} features have been observed 
in the diametrically opposite edge, where a strong X-ray continuum is also 
observed. 

The bipolar symmetry of the regions emitting this strong X-ray continuum may 
be an indication of particle acceleration with an external magnetic field 
tangential to the shock front in these regions as in the case of SN~1006 
(Reynolds 1996, 1998).

Fig.~4 (bottom panel) shows the MOS~1 and MOS~2 spectra of Knot A and
Knot B (highlighted in Fig.~5). The spectrum of Knot A features 
a strong high energy continuum, whereas the spectrum of Knot B is typical of 
the brightest silicon dominated regions.  We have used a single 
non-equilibrium model to fit the data (VNEI, XSPEC). The silicon and 
sulphur abundances in Knot A are close to solar, but all other elements have 
much lower values, in particular iron (see Table 1). The temperature is more 
than a factor 2 higher in Knot A (k$T = 2.3 \pm 0.3$~keV) for an ionization 
timescale ($n_{\mathrm e}t \simeq 1.4 \pm 0.4 \times 10^{10}$~s~cm$^{-3}$) 
three times shorter than in Knot B (k$T = 0.9 \pm 0.1$~keV, $n_{\mathrm e}t 
\simeq 4.7\pm 0.3 \times 10^{10}$~s~cm$^{-3}$). 

Both the narrow band images and the spectral analysis indicate that the 
continuum dominated emission most probably arises from the shocked ambient 
medium. We cannot, however, unambiguously identify the nature of this 
emission. The signature of a thermal plasma in this spectrum comes primarily 
from the silicon and sulphur lines, although they can also correspond to 
spatial contamination (due to the PSF) from the much brighter surrounding 
ejecta emission. A deeper systematic analysis of all the strong continuum 
knots is required to determine precisely their origin, which is beyond the 
scope of this paper.

\section {Conclusion}

We present the first results obtained from the {\it XMM-Newton} observation 
of the Tycho supernova remnant.\\
- The good correlation between the images in the \ion{Si}{xiii} K line and 
\ion{Fe}{xvii} L line implies that some fraction of the inner iron layer has 
been well mixed with the outer silicon layer.\\
- The fact that the Fe K line emission peaks distinctly at a smaller radius 
than the Fe L line emission is a clear indication that there is a spatial 
structure in the temperature in the ejecta, with higher temperature towards 
the reverse shock. This is coherent with the propagation of the reverse shock 
in the inner plateau of the ejecta.\\
- The silicon image correlates well at small scales with the radio image, and 
probably marks the contact discontinuity, distorted by Rayleigh-Taylor 
instabilities.\\
- The high-energy continuum, which is regular overall, peaks just behind the 
shock front, and is attributed to the emission from the shocked ambient 
medium.\\
- In addition to this regular continuum emission, we have discovered particularly 
bright and hard continuum knots on the eastern and western edges.

\begin{acknowledgements} 

We would like to warmly thank John Dickel for providing us with the 22 cm VLA 
radio image of Tycho as well as Philippe Gondoin for providing us with the 
PSF at different energies. 
M.~A. acknowledges support 
from the Swiss National Science Foundation (grants 2100-049343 and 
2000-058827), from the Swiss Academy of Sciences and from the Swiss 
Commission for Space Research.

\end{acknowledgements}

\end{document}